\documentclass[aps,prl,twocolumn,superscriptaddress,showpacs]{revtex4}

\usepackage{times,graphicx}

\newcommand{\apjl}{Astrophys. J.}

\newcommand{\jgr}{J. Geophys. Res.}
\newcommand{\apjs}{Astrophys. J.}
\newcommand{\ppcf}{Plasma Phys. Controlled Fusion}

\newcommand{\soa}{Soviet Astronomy}
\newcommand{\pof}{Phys. Fluids}
\newcommand{\pop}{Phys. Plasmas}
\newcommand{\lrsp}{Living Rev. Solar Phys.}
\newcommand{\ang}{Ann. Geophys.}
\newcommand{\ssr}{Space Sci. Rev.}
\newcommand{\jltp}{J. Low Temp. Phys.}

\newcommand{\rslpsa}{Proc. R. Soc. A}
\newcommand{\araa}{Annu. Rev. Astron. Astrophys.}

\newcommand{\Alfven}{Alfv\'en}
\newcommand{\Alfvenic}{Alfv\'enic}
\newcommand{\para}{\parallel}
\newcommand{\dBpara}{$\delta B^2_{\para}$}
\newcommand{\dBperp}{$\delta B^2_{\perp}$}

\begin{document}

\title{Anisotropy of Solar Wind Turbulence between Ion and Electron Scales}

\author{C.~H.~K.~Chen}
\email[]{christopher.chen03@imperial.ac.uk}
\author{T.~S.~Horbury}
\affiliation{The Blackett Laboratory, Imperial College London, London SW7 2AZ, United Kingdom}
\author{A.~A.~Schekochihin}
\affiliation{Rudolf Peierls Centre for Theoretical Physics, University of Oxford, Oxford OX1 3NP, United Kingdom}
\author{R.~T.~Wicks}
\affiliation{The Blackett Laboratory, Imperial College London, London SW7 2AZ, United Kingdom}
\author{O.~Alexandrova}
\affiliation{LESIA, Observatoire de Paris, CNRS, UPMC, Universit\'{e} Paris Diderot, 92190 Meudon, France}
\author{J.~Mitchell}
\affiliation{The Blackett Laboratory, Imperial College London, London SW7 2AZ, United Kingdom}

\begin{abstract}
The anisotropy of turbulence in the fast solar wind, between the ion and electron gyroscales, is directly observed using a multispacecraft analysis technique. Second order structure functions are calculated at different angles to the local magnetic field, for magnetic fluctuations both perpendicular and parallel to the mean field. In both components, the structure function value at large angles to the field $S_{\perp}$ is greater than at small angles $S_{\para}$: in the perpendicular component $S_{\perp}/S_{\para} = 5 \pm 1$  and in the parallel component $S_{\perp}/S_{\para} > 3$, implying spatially anisotropic fluctuations, $k_{\perp} > k_{\para}$. The spectral index of the perpendicular component is $-2.6$ at large angles and $-3$ at small angles, in broad agreement with critically balanced whistler and kinetic \Alfven\ wave predictions. For the parallel component, however, it is shallower than $-1.9$, which is considerably less steep than predicted for a kinetic \Alfven\ wave cascade.
\end{abstract}

\pacs{94.05.Lk, 52.35.Ra, 96.60.Vg, 96.50.Bh}

\maketitle

\emph{Introduction}.---Solar wind turbulence has been studied for many decades (e.g., \cite{goldsteinhorburybruno} and references therein) but a number of fundamental aspects of it remain poorly understood. This Letter will address one of these, the nature of the turbulent fluctuations at small scales, using a recently developed multispacecraft analysis technique.

Turbulence is usually modeled as a cascade of energy, with injection at large scales and dissipation at small scales. In the solar wind, the injected energy is thought to originate from the observed large scale \Alfven\ waves \cite{belcher71}. For scales between the effective outer scale and the ion gyroradius, termed the inertial range, a cascade of \Alfvenic\ fluctuations \cite{iroshnikov64,kraichnan65,goldreich95,boldyrev06,schekochihin09} is often invoked to explain the observed power spectra \cite{coleman68,matthaeus82a,bale05,podesta07}. One aspect of recent investigation in the solar wind inertial range, relevant to this study, is anisotropy with respect to the magnetic field. It has been shown that both power and scalings vary with respect to the local magnetic field direction \cite{bieber96,horbury08,podesta09a,osman09a,wicks10}, in a way consistent with critical balance theories \cite{goldreich95,boldyrev06}.

At smaller scales, close to the ion gyroradius, the magnetic field power spectrum steepens (e.g., \cite{leamon98a,smith06}). This is commonly termed the dissipation range, although is sometimes called the dispersion range (e.g., \cite{stawicki01}), and is where kinetic effects become important. Recent measurements of the magnetic field spectral index in this range are between $-2.3$ and $-2.8$ \cite{alexandrova08b,sahraoui09,kiyani09a,alexandrova09}, although larger variation was seen in an earlier survey \cite{smith06}. A further steepening in the spectrum near the electron gyroscale has also been observed \cite{sahraoui09,alexandrova09}. In this study, we investigate between the ion and electron scales. Two popular suggestions for the types of fluctuations in this range are kinetic \Alfven\ waves (KAWs) \cite{leamon98a,bale05,schekochihin09,sahraoui09,howes09} and whistler waves \cite{stawicki01,whistlers}. It has been suggested \cite{cho04,schekochihin09} that, like some inertial range theories \cite{goldreich95,boldyrev06}, the fluctuations are critically balanced, which would imply a spectral index of $-7/3$ in the perpendicular direction and $-5$ in the parallel direction.

In this Letter, the first multispacecraft structure function measurements in the solar wind at scales below the ion gyroscale are presented. The variance, power, and spectral index anisotropy in the magnetic field components parallel and perpendicular to the field are calculated. This provides a direct test of existing theories and a guide for new ones.

\emph{Data set}.---We use an interval of data from the Cluster mission \cite{escoubet01}, in which the four spacecraft are in the fast solar wind with a separation $\sim$100 km. The interval parameters are given in Table \ref{tab:parameters} and are from the FGM \cite{balogh01}, CIS \cite{reme01}, and PEACE \cite{johnstone97} instruments. No effects of Earth's foreshock are present, and the interval lies in the stable region of the parameter space for pressure anisotropy instabilities (e.g., \cite{bale09}).

\begin{table*}
\caption{\label{tab:parameters}Interval parameters ($V_{\text{SW}}$, solar wind speed; $n$, number density; $v_A$, \Alfven\ speed; $T$, temperature; $\rho$, gyroradius; $d$, inertial length; $\beta$, plasma beta)}
\begin{ruledtabular}
\begin{tabular}{cccccccccccccccccc}
Date & Time & $V_{\text{SW}}$ & $n_i$ & $v_A$ & $T_{\perp i}$ & $T_{\para i}$ & $T_{\perp e}$ & $T_{\para e}$ & $\rho_i$ & $\rho_e$ & $d_i$ & $d_e$ & $\beta_{i \para}$ & $\beta_{e \para}$ \\
(dd/mm/yy) & (UT) & (km s$^{-1}$) & (cm$^{-3}$) & (km s$^{-1}$) & (eV) & (eV) & (eV) & (eV) & (km) & (km) & (km) & (km) &  & \\
\hline
11/02/02 & 19:19--20:29 & $570$ & $3.8$ & $79$ & $23$ & $39$ & $12$ & $19$ & $94$ & $1.6$ & $120$ & $2.7$ & $1.1$ & $0.54$ \\
\end{tabular}
\end{ruledtabular}
\end{table*}

For analyzing the fluctuations between ion and electron scales, high frequency data, $>$1 Hz, is needed. In this study, a measurement of the local magnetic field direction is used, requiring data valid at both low and high frequencies. The STAFF instrument \cite{cornilleau-wehrlin03} has a high frequency search coil magnetometer, which provides a time series valid in the approximate range 0.6--10 Hz. We combine this with the FGM data, which is valid up to $\approx$1 Hz in the solar wind.

The combining procedure is based on the method in Appendix A of Ref.~\cite{alexandrova04}. First, the high frequency (22 Hz) FGM data are interpolated onto the times of the STAFF data (25 Hz). A wavelet transform is then applied to both time series to obtain two sets of wavelet coefficients for each field component. The wavelet coefficients corresponding to the STAFF data above 1 Hz are used to generate a high frequency time series, and those corresponding to the FGM data below 1 Hz are used to generate a low frequency time series. The two time series are then added, resulting in the combined signal.

\begin{figure}[b]
\includegraphics[scale=0.45]{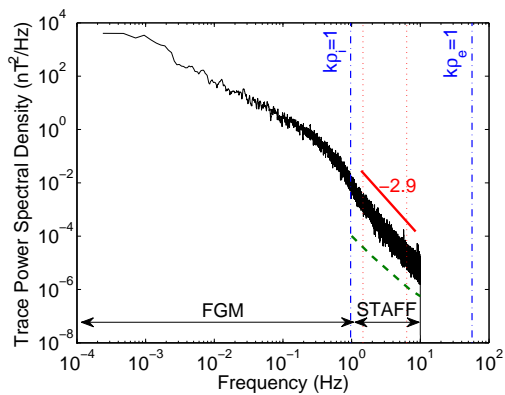}
\caption{\label{psd}(color online). Magnetic field power spectrum of combined STAFF and FGM data from Cluster 3. Approximate instrument ranges are shown, as is the STAFF noise floor (green dashed line) and ion and electron gyroscales (blue dash-dot lines). Approximate frequencies corresponding to the range of scales used are marked (red dotted lines).}
\end{figure}
  
The power spectrum of the combined data is shown in Fig.~\ref{psd} with the approximate ranges of FGM and STAFF marked. The noise floor of STAFF (from ground and in-flight tests \cite{cornilleau-wehrlin03}) is also shown. The break in the spectrum at $\approx$0.4 Hz is the ion scale spectral break point at the end of the inertial range, and is not due to the data merging. For this interval, the isotropic spectral index for the range of scales we study is $-2.88 \pm 0.01$.

\emph{Method}.---A multispacecraft method is used in which data from the four Cluster spacecraft are combined to produce second order structure functions in different directions to the local magnetic field. It is based on the method of Ref.~\cite{osman09a}. One benefit of this technique is that a range of sampling angles can be covered simultaneously, enabling short intervals to be used, increasing the likelihood of statistical stationarity.

Compared to the solar wind flow, the four spacecraft are approximately stationary and measure the magnetic field as the solar wind passes by. The measured variations, therefore, are due to both temporal and spatial variations in the plasma. Assuming that the temporal changes happen slowly compared to the flow, each time series can be converted into a spatial cut through the plasma (Taylor's hypothesis \cite{taylor38}). Second order structure functions can then be calculated, defined as $\delta B^2_i (\mathbf{l}) = \left\langle |B_i (\mathbf{r}+\mathbf{l})-B_i(\mathbf{r})|^2\right\rangle$, where $B_i$ is the $i$th component of the magnetic field, $\mathbf{l}$ is the separation vector, and the angular brackets denote an ensemble average over positions $\mathbf{r}$.

It is important to consider the application of Taylor's hypothesis at small scales. In the inertial range, the solar wind speed is usually an order of magnitude larger than the \Alfven\ speed and, therefore, Taylor's hypothesis is well satisfied. At smaller scales, the wave phase speed is larger than the \Alfven\ speed \cite{bale05,sahraoui09}. It is still lower than the solar wind speed, however, so even if Taylor's hypothesis is less well satisfied than in the inertial range, it is not an unreasonable assumption. Measurements with an \Alfvenic\ Taylor ratio (as defined in Ref.~\cite{osman09a}) greater than 0.25 are discarded.

Axisymmetry about the magnetic field is assumed so that $\mathbf{l}$ can be split into parallel and perpendicular components, $\mathbf{l}=(l_{\para},l_{\perp})$. There is mounting evidence that it is the \emph{local} magnetic field that orders the fluctuations rather than a \emph{global} field \cite{cho00,maron01,horbury08,beresnyak09,tessein09}; i.e., the turbulence is anisotropic with respect to the field at the scale of each fluctuation rather than a much larger scale. Here, the local field is defined as $\mathbf{B}_{local} = [\mathbf{B}(\mathbf{r}+\mathbf{l})+\mathbf{B}(\mathbf{r})]/2$, and its direction is used to define $l_{\para}$ and $l_{\perp}$. The parallel and perpendicular components of $\mathbf{B}$ for each structure function pair are also defined with respect to $\mathbf{B}_{local}$.

\begin{figure}[b]
\includegraphics[scale=0.67]{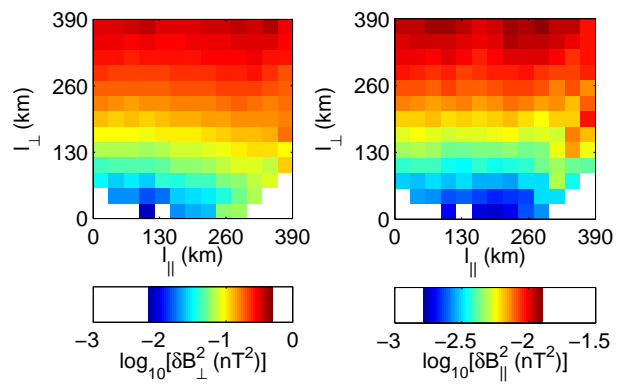}
\caption{\label{fig:S2PP}(color online). Second order structure functions with respect to spatial separations parallel ($l_{\para}$) and perpendicular ($l_{\perp}$) to the local magnetic field: \dBperp\ (left) and \dBpara\ (right).}
\end{figure}

The structure function values obtained from many pairs of magnetic field measurements from all four spacecraft are binned with respect to $l_{\para}$ and $l_{\perp}$ and averaged. A minimum number of 200 values per bin is set to ensure reliable results and the binned data for each component are shown in Fig.~\ref{fig:S2PP}. In both \dBperp\ and \dBpara\ anisotropy can be seen: the structure function contours are elongated along the local field direction. Similar results have been seen at larger scales in inertial range solar wind correlation functions \cite{osman07}, and in structure functions from MHD \cite{cho00} and electron MHD \cite{cho04,cho09} simulations. 

\begin{figure}
\includegraphics[scale=0.51]{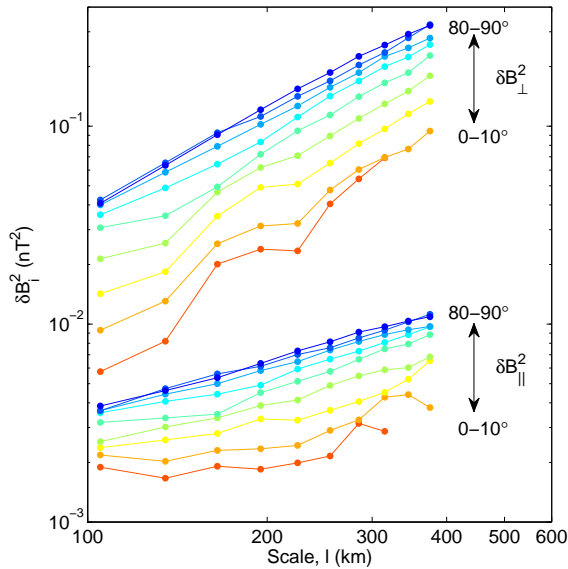}
\caption{\label{fig:S2AM}(color online). Second order structure functions at different angles to the local magnetic field for the perpendicular component (upper set) and the parallel component (lower set).}
\end{figure}

Instead of Cartesian coordinates, $(l_{\para},l_{\perp})$, the data can be binned in polar coordinates, $(l,\theta)$, where $\theta$ is the angle to the local magnetic field (see Fig.~\ref{fig:S2AM}). Because of the low power, it is possible that the noise floor of STAFF has been reached for the small angle bins in \dBpara. This can be seen in Fig.~\ref{fig:S2AM}, in which the lowest value structure function curves appear flatter than the others. Caution, therefore, is advised when interpreting these lowest power measurements. 

\emph{Variance anisotropy}.---The ratio of power in the perpendicular component to the parallel component is sometimes referred to as variance anisotropy, (e.g., \cite{hamilton08}). From Fig.~\ref{fig:S2AM} it can be seen that \dBpara\ is about 5\% of \dBperp, which is smaller than average values of previous measurements \cite{leamon98a,hamilton08}. This could be due to statistical variation, or due to the global, rather than local, mean field direction being used in those studies.

The variance anisotropy for KAWs in electron reduced MHD \cite{schekochihin09} can be calculated, and for the parameters in Table \ref{tab:parameters} this prediction is $\delta B^2_{\para}/\delta B^2_{\perp}=0.4$, which is larger than the value observed here. Numerical solutions of linear kinetic theory, however, suggest smaller values of variance anisotropy for KAWs \cite{gary09}. These values depend on $\beta$, propagation angle, and wave number, but are in the range 0.01 to 0.2, which agrees with our result of 0.05.

\begin{figure}
\includegraphics[scale=0.53]{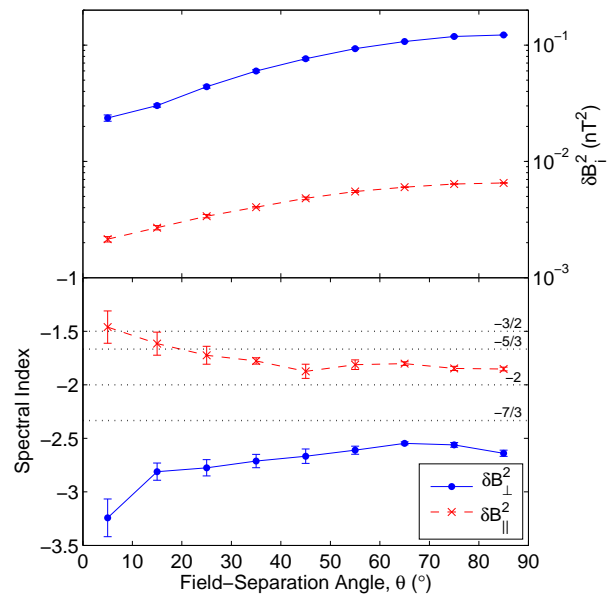}
\caption{\label{fig:anisotropy}(color online). Anisotropy of second order structure functions (upper) and spectral indices (lower). Various spectral index predictions are marked (dotted lines).}
\end{figure}

\emph{Power anisotropy}.---Straight lines (in log-log space) are fitted to the data in Fig.~\ref{fig:S2AM} over the range 100--400 km. This is between ion and electron scales, i.e., between $k\rho_i=1$ and $k\rho_e=1$, where $k=2\pi/l$. The interpolated values of the structure function at 200 km from these fits are given as a function of $\theta$ in the upper panel of Fig.~\ref{fig:anisotropy}. The error bars are small, comparable to the data point size, and are the standard deviations of the points about the best fit lines.

For both components, the structure function value (``power'') increases with $\theta$. This is consistent with spatially anisotropic fluctuations, $k_{\perp}>k_{\para}$, where $k_{\perp}$ and $k_{\para}$ are characteristic parallel and perpendicular wave numbers \cite{chen10a}. One measure of this anisotropy is the ratio of the largest angle bin value $S_{\perp}$ to the smallest $S_{\para}$, which for \dBperp\ is $S_{\perp}/S_{\para} = 5 \pm 1$. This number is uncertain for \dBpara, due to the noise issues mentioned above, but has a lower limit of 3.

Some previous studies of the solar wind between ion and electron scales have measured anisotropy using the ``slab plus 2D'' model \cite{leamon98a,hamilton08}. The large slab fractions obtained are not generally in agreement with this study. In fact, our results are more consistent with other solar wind \cite{narita07a,podesta09a} and magnetosheath \cite{alexandrova08c} measurements that demonstrate significant spatial anisotropy at scales smaller than the ion gyroscale.

\emph{Spectral index anisotropy}.---An important characteristic of turbulence is the spectral index $-\alpha$ of the power spectrum, $E(k) \sim k^{-\alpha}$. Second order structure function scalings, i.e., gradients $g$, of the straight line fits to the data in Fig.~\ref{fig:S2AM}, are related to the spectral index by $\alpha=g+1$ \cite{monin75}. Using this relationship, the spectral index as a function of angle $\theta$, for \dBperp\ and \dBpara , is shown in Fig.~\ref{fig:anisotropy}. The error bars are the standard errors on the best fit line gradients.

For \dBperp, the spectral index varies from around $-2.6$ at large angles to $-3.2$ at small angles. It should be noted that the steepest spectral index it is possible to measure with this method is $-3$ (e.g., \cite{monin75,cho09}); for steeper spectra, the scaling seen by the two-point second order structure function is $g=2$, since it is dominated by the smooth variation of the large scale field. At small angles, we observe a spectral index of $-3.2 \pm 0.2$, indicating that the spectrum in the parallel direction is $k_{\para}^{-3}$ or steeper. The predictions for a critically balanced whistler or KAW cascade are $-7/3$ in the perpendicular direction and $-5$ in the parallel direction \cite{cho04,schekochihin09}. Although the spectral indices in Fig.~\ref{fig:anisotropy} are slightly steeper than the prediction at large values of $\theta$, the steepening towards small $\theta$ is suggestive of a critically balanced cascade.

The spectral index of \dBpara\ varies from $-1.9$ at large angles to $-1.5$ at small angles. The small angle values may be affected by noise (as discussed previously), but the large angle ones appear not to be, and are significantly shallower than those of \dBperp. This difference in gradient between the components can also be seen in Fig.~\ref{fig:S2AM}. For a KAW cascade, \dBpara\ is expected to scale in the same way as \dBperp\ \cite{schekochihin09}. The difference observed here, therefore, may be indicating the presence of other modes or a different cascade mechanism. Another possibility for the difference is instability generated fluctuations, although the measured parameters suggest the interval is not unstable to pressure anisotropy instabilities (e.g., \cite{bale09}).

\emph{Summary and conclusions}.---The variance, power, and spectral index anisotropy are measured in the fast solar wind, between the ion and electron gyroscales. The variance anisotropy is significant, with \dBpara\ being approximately 5\% of \dBperp. Both magnetic field components display power anisotropy, implying spatially anisotropic fluctuations, $k_{\perp}>k_{\para}$. The spectral index of \dBperp\ steepens at small angles to the field, which is consistent with a critically balanced cascade of whistlers or KAWs. The spectral indices of \dBpara are less consistent with the predictions, suggesting that the KAW picture \cite{schekochihin09} may be incomplete.

Although we have looked for other data intervals, it is hard to find ones that satisfy the conditions required for this analysis, i.e., $\sim$1 h long, away from Earth's foreshock, with small spacecraft separations and good angular coverage. A larger study is required to determine if the behavior noted here is typical for the solar wind. This may need to wait for a future mission due to the limitations of multispacecraft data currently available.

\begin{acknowledgments}
This work was funded by STFC and the Leverhulme Trust Network for Magnetized Plasma Turbulence. FGM and CIS data were obtained from the Cluster Active Archive. C.~C.~acknowledges useful conversations with K.~Osman, P.~Brown, and S.~Schwartz.
\end{acknowledgments}


\begin{thebibliography}{52}
\expandafter\ifx\csname natexlab\endcsname\relax\def\natexlab#1{#1}\fi
\expandafter\ifx\csname bibnamefont\endcsname\relax
  \def\bibnamefont#1{#1}\fi
\expandafter\ifx\csname bibfnamefont\endcsname\relax
  \def\bibfnamefont#1{#1}\fi
\expandafter\ifx\csname citenamefont\endcsname\relax
  \def\citenamefont#1{#1}\fi
\expandafter\ifx\csname url\endcsname\relax
  \def\url#1{\texttt{#1}}\fi
\expandafter\ifx\csname urlprefix\endcsname\relax\def\urlprefix{URL }\fi
\providecommand{\bibinfo}[2]{#2}
\providecommand{\eprint}[2][]{\url{#2}}

\bibitem[{\citenamefont{Goldstein\emph{\ et\ al.}}(1995)}]{goldsteinhorburybruno}
\bibinfo{author}{\bibfnamefont{M.~L.} \bibnamefont{Goldstein\emph{\ et\ al.}}},
  \bibinfo{journal}{\araa} \textbf{\bibinfo{volume}{33}}, \bibinfo{pages}{283}
  (\bibinfo{year}{1995});
%\bibitem[{\citenamefont{Horbury\emph{\ et\ al.}}(2005)}]{horbury05}
\bibinfo{author}{\bibfnamefont{T.~S.} \bibnamefont{Horbury\emph{\ et\ al.}}},
  \bibinfo{journal}{\ppcf} \textbf{\bibinfo{volume}{47}}, \bibinfo{pages}{B703}
  (\bibinfo{year}{2005});
%\bibitem[{\citenamefont{{Bruno} and {Carbone}}(2005)}]{bruno05}
\bibinfo{author}{\bibfnamefont{R.}~\bibnamefont{{Bruno}}} \bibnamefont{and}
  \bibinfo{author}{\bibfnamefont{V.}~\bibnamefont{{Carbone}}},
  \bibinfo{journal}{\lrsp} \textbf{\bibinfo{volume}{2}}, \bibinfo{pages}{4}
  (\bibinfo{year}{2005}).

\bibitem[{\citenamefont{{Belcher} and {Davis}}(1971)}]{belcher71}
\bibinfo{author}{\bibfnamefont{J.~W.} \bibnamefont{{Belcher}}}
  \bibnamefont{and} \bibinfo{author}{\bibfnamefont{L.}~\bibnamefont{{Davis}},
  \bibfnamefont{Jr.}}, \bibinfo{journal}{\jgr} \textbf{\bibinfo{volume}{76}},
  \bibinfo{pages}{3534} (\bibinfo{year}{1971}).

\bibitem[{\citenamefont{{Iroshnikov}}(1964)}]{iroshnikov64}
\bibinfo{author}{\bibfnamefont{P.~S.} \bibnamefont{{Iroshnikov}}},
  \bibinfo{journal}{\soa} \textbf{\bibinfo{volume}{7}}, \bibinfo{pages}{566}
  (\bibinfo{year}{1964}).

\bibitem[{\citenamefont{{Kraichnan}}(1965)}]{kraichnan65}
\bibinfo{author}{\bibfnamefont{R.~H.} \bibnamefont{{Kraichnan}}},
  \bibinfo{journal}{\pof} \textbf{\bibinfo{volume}{8}}, \bibinfo{pages}{1385}
  (\bibinfo{year}{1965}).

\bibitem[{\citenamefont{{Goldreich} and {Sridhar}}(1995)}]{goldreich95}
\bibinfo{author}{\bibfnamefont{P.}~\bibnamefont{{Goldreich}}} \bibnamefont{and}
  \bibinfo{author}{\bibfnamefont{S.}~\bibnamefont{{Sridhar}}},
  \bibinfo{journal}{\apj} \textbf{\bibinfo{volume}{438}}, \bibinfo{pages}{763}
  (\bibinfo{year}{1995}).

\bibitem[{\citenamefont{{Boldyrev}}(2006)}]{boldyrev06}
\bibinfo{author}{\bibfnamefont{S.}~\bibnamefont{{Boldyrev}}},
  \bibinfo{journal}{\prl} \textbf{\bibinfo{volume}{96}},
  \bibinfo{pages}{115002} (\bibinfo{year}{2006}).

\bibitem[{\citenamefont{Schekochihin\emph{\ et\ al.}}(2009)}]{schekochihin09}
\bibinfo{author}{\bibfnamefont{A.~A.} \bibnamefont{Schekochihin\emph{\ et\
  al.}}}, \bibinfo{journal}{\apjs} \textbf{\bibinfo{volume}{182}},
  \bibinfo{pages}{310} (\bibinfo{year}{2009}).

\bibitem[{\citenamefont{{Coleman}}(1968)}]{coleman68}
\bibinfo{author}{\bibfnamefont{P.~J.} \bibnamefont{{Coleman}},
  \bibfnamefont{Jr.}}, \bibinfo{journal}{\apj} \textbf{\bibinfo{volume}{153}},
  \bibinfo{pages}{371} (\bibinfo{year}{1968}).

\bibitem[{\citenamefont{{Matthaeus} and {Goldstein}}(1982)}]{matthaeus82a}
\bibinfo{author}{\bibfnamefont{W.~H.} \bibnamefont{{Matthaeus}}}
  \bibnamefont{and} \bibinfo{author}{\bibfnamefont{M.~L.}
  \bibnamefont{{Goldstein}}}, \bibinfo{journal}{\jgr}
  \textbf{\bibinfo{volume}{87}}, \bibinfo{pages}{6011} (\bibinfo{year}{1982}).

\bibitem[{\citenamefont{Bale\emph{\ et\ al.}}(2005)}]{bale05}
\bibinfo{author}{\bibfnamefont{S.~D.} \bibnamefont{Bale\emph{\ et\ al.}}},
  \bibinfo{journal}{\prl} \textbf{\bibinfo{volume}{94}},
  \bibinfo{pages}{215002} (\bibinfo{year}{2005}).

\bibitem[{\citenamefont{Podesta\emph{\ et\ al.}}(2007)}]{podesta07}
\bibinfo{author}{\bibfnamefont{J.~J.} \bibnamefont{Podesta\emph{\ et\ al.}}},
  \bibinfo{journal}{\apj} \textbf{\bibinfo{volume}{664}}, \bibinfo{pages}{543}
  (\bibinfo{year}{2007}).

\bibitem[{\citenamefont{Bieber\emph{\ et\ al.}}(1996)}]{bieber96}
\bibinfo{author}{\bibfnamefont{J.~W.} \bibnamefont{Bieber\emph{\ et\ al.}}},
  \bibinfo{journal}{\jgr} \textbf{\bibinfo{volume}{101}}, \bibinfo{pages}{2511}
  (\bibinfo{year}{1996}).

\bibitem[{\citenamefont{Horbury\emph{\ et\ al.}}(2008)}]{horbury08}
\bibinfo{author}{\bibfnamefont{T.~S.} \bibnamefont{Horbury\emph{\ et\ al.}}},
  \bibinfo{journal}{\prl} \textbf{\bibinfo{volume}{101}},
  \bibinfo{pages}{175005} (\bibinfo{year}{2008}).

\bibitem[{\citenamefont{{Podesta}}(2009)}]{podesta09a}
\bibinfo{author}{\bibfnamefont{J.~J.} \bibnamefont{{Podesta}}},
  \bibinfo{journal}{\apj} \textbf{\bibinfo{volume}{698}}, \bibinfo{pages}{986}
  (\bibinfo{year}{2009}).

\bibitem[{\citenamefont{{Osman} and {Horbury}}(2009)}]{osman09a}
\bibinfo{author}{\bibfnamefont{K.~T.} \bibnamefont{{Osman}}} \bibnamefont{and}
  \bibinfo{author}{\bibfnamefont{T.~S.} \bibnamefont{{Horbury}}},
  \bibinfo{journal}{\ang} \textbf{\bibinfo{volume}{27}}, \bibinfo{pages}{3019}
  (\bibinfo{year}{2009}).

\bibitem[{\citenamefont{Wicks\emph{\ et\ al.}}(2010)}]{wicks10}
\bibinfo{author}{\bibfnamefont{R.~T.} \bibnamefont{Wicks\emph{\ et\ al.}}},
  \bibinfo{journal}{arXiv:1002.2096v1}.

\bibitem[{\citenamefont{Leamon\emph{\ et\ al.}}(1998)}]{leamon98a}
\bibinfo{author}{\bibfnamefont{R.~J.} \bibnamefont{Leamon\emph{\ et\ al.}}},
  \bibinfo{journal}{\jgr} \textbf{\bibinfo{volume}{103}}, \bibinfo{pages}{4775}
  (\bibinfo{year}{1998}).

\bibitem[{\citenamefont{Smith\emph{\ et\ al.}}(2006)}]{smith06}
\bibinfo{author}{\bibfnamefont{C.~W.} \bibnamefont{Smith\emph{\ et\ al.}}},
  \bibinfo{journal}{\apjl} \textbf{\bibinfo{volume}{645}}, \bibinfo{pages}{L85}
  (\bibinfo{year}{2006}).

\bibitem[{\citenamefont{Stawicki\emph{\ et\ al.}}(2001)}]{stawicki01}
\bibinfo{author}{\bibfnamefont{O.}~\bibnamefont{Stawicki\emph{\ et\ al.}}},
  \bibinfo{journal}{\jgr} \textbf{\bibinfo{volume}{106}}, \bibinfo{pages}{8273}
  (\bibinfo{year}{2001}).

\bibitem[{\citenamefont{Alexandrova\emph{\ et\
  al.}}(2008{\natexlab{a}})}]{alexandrova08b}
\bibinfo{author}{\bibfnamefont{O.}~\bibnamefont{Alexandrova\emph{\ et\ al.}}},
  \bibinfo{journal}{\apj} \textbf{\bibinfo{volume}{674}}, \bibinfo{pages}{1153}
  (\bibinfo{year}{2008}{\natexlab{a}}).

\bibitem[{\citenamefont{Sahraoui\emph{\ et\ al.}}(2009)}]{sahraoui09}
\bibinfo{author}{\bibfnamefont{F.}~\bibnamefont{Sahraoui\emph{\ et\ al.}}},
  \bibinfo{journal}{\prl} \textbf{\bibinfo{volume}{102}},
  \bibinfo{pages}{231102} (\bibinfo{year}{2009}).

\bibitem[{\citenamefont{Kiyani\emph{\ et\ al.}}(2009)}]{kiyani09a}
\bibinfo{author}{\bibfnamefont{K.~H.} \bibnamefont{Kiyani\emph{\ et\ al.}}},
  \bibinfo{journal}{\prl} \textbf{\bibinfo{volume}{103}},
  \bibinfo{pages}{075006} (\bibinfo{year}{2009}).

\bibitem[{\citenamefont{Alexandrova\emph{\ et\ al.}}(2009)}]{alexandrova09}
\bibinfo{author}{\bibfnamefont{O.}~\bibnamefont{Alexandrova\emph{\ et\ al.}}},
  \bibinfo{journal}{\prl} \textbf{\bibinfo{volume}{103}},
  \bibinfo{pages}{165003} (\bibinfo{year}{2009}).

\bibitem[{\citenamefont{{Howes} and {Quataert}}(2010)}]{howes09}
\bibinfo{author}{\bibfnamefont{G.~G.} \bibnamefont{{Howes}}} \bibnamefont{and}
  \bibinfo{author}{\bibfnamefont{E.}~\bibnamefont{{Quataert}}},
  \bibinfo{journal}{\apjl} \textbf{\bibinfo{volume}{709}}, \bibinfo{pages}{L49}
  (\bibinfo{year}{2010}).

\bibitem[{\citenamefont{Biskamp\emph{\ et\ al.}}(1996)}]{whistlers}
\bibinfo{author}{\bibfnamefont{D.}~\bibnamefont{Biskamp\emph{\ et\ al.}}},
  \bibinfo{journal}{\prl} \textbf{\bibinfo{volume}{76}}, \bibinfo{pages}{1264}
  (\bibinfo{year}{1996});
%\bibitem[{\citenamefont{{Galtier}}(2006)}]{galtier06}
\bibinfo{author}{\bibfnamefont{S.}~\bibnamefont{{Galtier}}},
  \bibinfo{journal}{\jltp} \textbf{\bibinfo{volume}{145}}, \bibinfo{pages}{59}
  (\bibinfo{year}{2006});
%\bibitem[{\citenamefont{Matthaeus\emph{\ et\ al.}}(2008)}]{matthaeus08}
\bibinfo{author}{\bibfnamefont{W.~H.} \bibnamefont{Matthaeus\emph{\ et\ al.}}},
  \bibinfo{journal}{\prl} \textbf{\bibinfo{volume}{101}},
  \bibinfo{pages}{149501} (\bibinfo{year}{2008});
%\bibitem[{\citenamefont{Saito\emph{\ et\ al.}}(2008)}]{saito08}
\bibinfo{author}{\bibfnamefont{S.}~\bibnamefont{Saito\emph{\ et\ al.}}},
  \bibinfo{journal}{\pop} \textbf{\bibinfo{volume}{15}},
  \bibinfo{pages}{102305} (\bibinfo{year}{2008}).

\bibitem[{\citenamefont{{Cho} and {Lazarian}}(2004)}]{cho04}
\bibinfo{author}{\bibfnamefont{J.}~\bibnamefont{{Cho}}} \bibnamefont{and}
  \bibinfo{author}{\bibfnamefont{A.}~\bibnamefont{{Lazarian}}},
  \bibinfo{journal}{\apjl} \textbf{\bibinfo{volume}{615}}, \bibinfo{pages}{L41}
  (\bibinfo{year}{2004}).

\bibitem[{\citenamefont{Escoubet\emph{\ et\ al.}}(2001)}]{escoubet01}
\bibinfo{author}{\bibfnamefont{C.~P.} \bibnamefont{Escoubet\emph{\ et\ al.}}},
  \bibinfo{journal}{\ang} \textbf{\bibinfo{volume}{19}}, \bibinfo{pages}{1197}
  (\bibinfo{year}{2001}).

\bibitem[{\citenamefont{Balogh\emph{\ et\ al.}}(2001)}]{balogh01}
\bibinfo{author}{\bibfnamefont{A.}~\bibnamefont{Balogh\emph{\ et\ al.}}},
  \bibinfo{journal}{\ang} \textbf{\bibinfo{volume}{19}}, \bibinfo{pages}{1207}
  (\bibinfo{year}{2001}).

\bibitem[{\citenamefont{{R{\`e}me}\emph{\ et\ al.}}(2001)}]{reme01}
\bibinfo{author}{\bibfnamefont{H.}~\bibnamefont{{R{\`e}me}\emph{\ et\ al.}}},
  \bibinfo{journal}{\ang} \textbf{\bibinfo{volume}{19}}, \bibinfo{pages}{1303}
  (\bibinfo{year}{2001}).

\bibitem[{\citenamefont{Johnstone\emph{\ et\ al.}}(1997)}]{johnstone97}
\bibinfo{author}{\bibfnamefont{A.~D.} \bibnamefont{Johnstone\emph{\ et\ al.}}},
  \bibinfo{journal}{\ssr} \textbf{\bibinfo{volume}{79}}, \bibinfo{pages}{351}
  (\bibinfo{year}{1997}).

\bibitem[{\citenamefont{Bale\emph{\ et\ al.}}(2009)}]{bale09}
\bibinfo{author}{\bibfnamefont{S.~D.} \bibnamefont{Bale\emph{\ et\ al.}}},
  \bibinfo{journal}{\prl} \textbf{\bibinfo{volume}{103}},
  \bibinfo{pages}{211101} (\bibinfo{year}{2009}).

\bibitem[{\citenamefont{Cornilleau-Wehrlin\emph{\ et\
  al.}}(2003)}]{cornilleau-wehrlin03}
\bibinfo{author}{\bibfnamefont{N.}~\bibnamefont{Cornilleau-Wehrlin\emph{\ et\
  al.}}}, \bibinfo{journal}{\ang} \textbf{\bibinfo{volume}{21}},
  \bibinfo{pages}{437} (\bibinfo{year}{2003}).

\bibitem[{\citenamefont{Alexandrova\emph{\ et\ al.}}(2004)}]{alexandrova04}
\bibinfo{author}{\bibfnamefont{O.}~\bibnamefont{Alexandrova\emph{\ et\ al.}}},
  \bibinfo{journal}{\jgr} \textbf{\bibinfo{volume}{109}}, \bibinfo{pages}{A05207}
  (\bibinfo{year}{2004}).

\bibitem[{\citenamefont{{Taylor}}(1938)}]{taylor38}
\bibinfo{author}{\bibfnamefont{G.~I.} \bibnamefont{{Taylor}}},
  \bibinfo{journal}{\rslpsa} \textbf{\bibinfo{volume}{164}},
  \bibinfo{pages}{476} (\bibinfo{year}{1938}).

\bibitem[{\citenamefont{{Cho} and {Vishniac}}(2000)}]{cho00}
\bibinfo{author}{\bibfnamefont{J.}~\bibnamefont{{Cho}}} \bibnamefont{and}
  \bibinfo{author}{\bibfnamefont{E.~T.} \bibnamefont{{Vishniac}}},
  \bibinfo{journal}{\apj} \textbf{\bibinfo{volume}{539}}, \bibinfo{pages}{273}
  (\bibinfo{year}{2000}).

\bibitem[{\citenamefont{{Maron} and {Goldreich}}(2001)}]{maron01}
\bibinfo{author}{\bibfnamefont{J.}~\bibnamefont{{Maron}}} \bibnamefont{and}
  \bibinfo{author}{\bibfnamefont{P.}~\bibnamefont{{Goldreich}}},
  \bibinfo{journal}{\apj} \textbf{\bibinfo{volume}{554}}, \bibinfo{pages}{1175}
  (\bibinfo{year}{2001}).

\bibitem[{\citenamefont{{Beresnyak} and {Lazarian}}(2009)}]{beresnyak09}
\bibinfo{author}{\bibfnamefont{A.}~\bibnamefont{{Beresnyak}}} \bibnamefont{and}
  \bibinfo{author}{\bibfnamefont{A.}~\bibnamefont{{Lazarian}}},
  \bibinfo{journal}{\apj} \textbf{\bibinfo{volume}{702}}, \bibinfo{pages}{460}
  (\bibinfo{year}{2009}).

\bibitem[{\citenamefont{Tessein\emph{\ et\ al.}}(2009)}]{tessein09}
\bibinfo{author}{\bibfnamefont{J.~A.} \bibnamefont{Tessein\emph{\ et\ al.}}},
  \bibinfo{journal}{\apj} \textbf{\bibinfo{volume}{692}}, \bibinfo{pages}{684}
  (\bibinfo{year}{2009}).

\bibitem[{\citenamefont{{Osman} and {Horbury}}(2007)}]{osman07}
\bibinfo{author}{\bibfnamefont{K.~T.} \bibnamefont{{Osman}}} \bibnamefont{and}
  \bibinfo{author}{\bibfnamefont{T.~S.} \bibnamefont{{Horbury}}},
  \bibinfo{journal}{\apjl} \textbf{\bibinfo{volume}{654}},
  \bibinfo{pages}{L103} (\bibinfo{year}{2007}).

\bibitem[{\citenamefont{{Cho} and {Lazarian}}(2009)}]{cho09}
\bibinfo{author}{\bibfnamefont{J.}~\bibnamefont{{Cho}}} \bibnamefont{and}
  \bibinfo{author}{\bibfnamefont{A.}~\bibnamefont{{Lazarian}}},
  \bibinfo{journal}{\apj} \textbf{\bibinfo{volume}{701}}, \bibinfo{pages}{236}
  (\bibinfo{year}{2009}).

\bibitem[{\citenamefont{Hamilton\emph{\ et\ al.}}(2008)}]{hamilton08}
\bibinfo{author}{\bibfnamefont{K.}~\bibnamefont{Hamilton\emph{\ et\ al.}}},
  \bibinfo{journal}{\jgr} \textbf{\bibinfo{volume}{113}}, \bibinfo{pages}{A01106}
  (\bibinfo{year}{2008}).

\bibitem[{\citenamefont{{Gary} and {Smith}}(2009)}]{gary09}
\bibinfo{author}{\bibfnamefont{S.~P.} \bibnamefont{{Gary}}} \bibnamefont{and}
  \bibinfo{author}{\bibfnamefont{C.~W.} \bibnamefont{{Smith}}},
  \bibinfo{journal}{\jgr} \textbf{\bibinfo{volume}{114}},
  \bibinfo{pages}{A12105} (\bibinfo{year}{2009}).

\bibitem[{\citenamefont{Chen\emph{\ et\ al.}}(2010)}]{chen10a}
\bibinfo{author}{\bibfnamefont{C.~H.~K.} \bibnamefont{Chen\emph{\ et\ al.}}},
  \bibinfo{journal}{\apjl} \textbf{\bibinfo{volume}{711}}, \bibinfo{pages}{L79}
  (\bibinfo{year}{2010}).

\bibitem[{\citenamefont{Narita\emph{\ et\ al.}}(2007)}]{narita07a}
\bibinfo{author}{\bibfnamefont{Y.}~\bibnamefont{Narita\emph{\ et\ al.}}}, in
  \emph{\bibinfo{booktitle}{Proceedings of the 6th Annual International Astrophysical Conference, Oahu, Hawaii, 2007}}, edited by
  \bibinfo{editor}{\bibfnamefont{D.}~\bibnamefont{{Shaikh}}} \bibnamefont{and}
  \bibinfo{editor}{\bibfnamefont{G.~P.} \bibnamefont{{Zank}}}
  (\bibinfo{publisher}{AIP, New York}, \bibinfo{year}{2007}), Vol.
  \bibinfo{volume}{932}, pp. \bibinfo{pages}{215-220}.

\bibitem[{\citenamefont{Alexandrova\emph{\ et\
  al.}}(2008{\natexlab{b}})}]{alexandrova08c}
\bibinfo{author}{\bibfnamefont{O.}~\bibnamefont{Alexandrova\emph{\ et\ al.}}},
  \bibinfo{journal}{\ang} \textbf{\bibinfo{volume}{26}}, \bibinfo{pages}{3585}
  (\bibinfo{year}{2008}{\natexlab{b}}).

\bibitem[{\citenamefont{{Monin} and {Yaglom}}(1975)}]{monin75}
\bibinfo{author}{\bibfnamefont{A.~S.} \bibnamefont{{Monin}}} \bibnamefont{and}
  \bibinfo{author}{\bibfnamefont{A.~M.} \bibnamefont{{Yaglom}}},
  \emph{\bibinfo{title}{{Statistical Fluid Mechanics, Vol 2}}}
  (\bibinfo{publisher}{MIT Press, Cambridge, Mass.}, \bibinfo{year}{1975}).

\end{thebibliography}
\end{document}